\documentclass[smallextended]{svjour3}
\smartqed  
\usepackage[utf8]{inputenc}
\usepackage{graphics}
\usepackage{epstopdf}
\usepackage{longtable}
\usepackage{pdfpages}
\usepackage{pgfplotstable}
\usepackage{amsmath}
\usepackage{amssymb}
\usepackage{textcomp}
\DeclareMathOperator{\diverg}{div}

\begin{document}
\title{Waves in strong centrifugal filed: dissipative gas}

\author{S.V. Bogovalov, V.A. Kislov,  I.V. Tronin}

\institute{I.V. Tronin \at
              National Research Nuclear University ``MEPhI'' Kashirskoye shosse, 31, Moscow, 115409, Russia \\
              Tel.: +7-495-788-5699 (9850)\\
              Fax: +7-495-324-21-11\\
              \email{IVTronin@mephi.ru}
}

\date{Received: date / Accepted: date}

\maketitle
\begin{abstract}
In the fast rotating gas (with the velocity typical for Iguassu gas centrifuge) three families of linear waves exist with different polarizations and law of dispersion. The energy of the waves is basically concentrated at the axis of rotation in the rarefied region. Therefore these waves decay on the distance comparable with the wavelength. There is only one type of waves propagating strictly along the axis of rotation with the law of dispersion similar to ordinary acoustic waves. These waves are interested for the physics of gas centrifuges. The energy density of these waves  concentrates at the wall of the rotor. These waves have weak damping due to the molecular viscosity and heat conductivity. The damping coefficient  is determined for this type of waves by numerical calculations. Analytical approximations for the damping coefficient is defined as well. At the parameters typical for the Iguassu centrifuge the damping is defined by interaction of the waves with the rotor wall. 

\keywords{High-speed flow \and gas dynamics \and general fluid mechanics \and rotating flows \and waves in rotating fluids}
\end{abstract}

\maketitle

\section{Introduction}
Wave dynamics in rotating liquids and gases attracts attention for a long time. Although the waves in the rotating liquids were studied in many works~\cite{GC:baines_1967,GC:greenspan,GC:kobine_1995}, the properties of the waves in the gases were studied only in few works and the investigations have been limited by rather moderate rotational velocities~\cite{GC:waves_2005}. Properties of these waves are interesting not only from the theoretical point of view rather than for practical use, in particular in physics of gas centrifuges (GC).

GC are explored for uranium isotope production starting with 50th years of the last century. In spite of this, the physics of the gas flow in the GC is still not fully understood. In our previous paper~\cite{GC:waves} we argued that the scoops used for the expulsion of the gas from the GC produce waves which propagate along the rotor. The properties of the waves in ideal gas compressed in the centrifugal field of the order $10^6 ~\rm g$ ($\rm g$ -- acceleration of gravity at the Earth surface) are rather specific.  

The velocities of the conventional GC is about $600-700~ \rm m/s$ at the rotor radius about $6-9~\rm cm$~\cite{GC:Glaser}. The centrifugal acceleration is about $10^5-10^6 ~\rm  g$ at these parameters. Therefore, pressure of the working gas changes on 3-4 orders of magnitude at $1 ~\rm cm$ radius variation. In these conditions conventional acoustic waves split into three families having different dispersion and polarization~\cite{GC:waves}.

In the previous work~\cite{GC:waves} our analysis has been limited to the waves in ideal dissipationless gas. The molecular viscosity and thermal conductivity have been neglected. This approach allowed us to simplify the problem and to obtain dispersion characteristics of all families of the waves. However, the damping of the waves due to the dissipation processes remains unknown. The solution of this problem is important for the physics of the gas flow in the gas centrifuges on a few reasons.

{\bf One of the key element of the mechanism of the centrifugal isotope separation  is the secondary axial circulation of the gas. This circulation dramatically multiplies the radial separative effect in the GC~\cite{GC:concurrent}.  In particular, scoops for extraction of the gas produce this circulation. The exitation of the axial circulation at the axially symmetric braking of the gas  by the end cups of the rotor has been investigated in the work ~\cite{GC:Wood_waves} assuming the steady state flow. The end cups imitate the impact of the scoops on the flow. The impact decays along the axis exponentially. But in more realistic picture the scoops produce waves along the axis. The  waves can produce an additional axial circulation  due to the  so-called ``acoustic flows'' which was discussed starting with the works by Lord Rayleigh~\cite{GC:Rayleigh1,GC:Rayleigh2}. Direct numerical experiments~\cite{GC:impact_waves} have shown that periodical braking of the gas by the scoops produce circulation which esentially differ from the circulation exited by the stationary braking of the gas.} The impact of the waves on the acoustic flow essentially depends on the rate of decay (damping) of the waves due to the dissipation. If the damping is strong, then the waves transfer their energy and momentum to the gas close to the scoops. If the damping is weak, the waves can propagate for a long distance from the scoops and transfer energy and momentum to the gas far from them. The axial circulation and multiplication of the radial separative effect can differ in these cases.

Additional motivation to study damping of the waves in GC is the impact of the waves on the gas content. Numerical experiments~\cite{GC:impact_aip} have shown that the waves change the gas flux from the waste chamber of GC provided that the waves can reach the opposite end of the rotor and product baffle. For the gas centrifuges this may result into reduction of the gas content in the gas centrifuges. This important effect depends on the damping of the waves as well.

The waves with the weakest damping are especially interesting for us because they can strongly modify the impact of the scoops on the axial circulation and gas content in the GC. In this paper we concentrate on the physics of damping of this type of waves. These waves have the dispersion law similar to the dispersion law of the conventional acoustic waves. They are polarized along the rotational axis and energy of these waves concentrates at the rotor wall.

The paper is organized as follows. In the second section we classify the wave families in rotating gas. The third and the fourth sections describe numerical method used throughout the paper and its verification. Next two sections consist of the results of the numerical solution of the problem and analytical estimations. In the section 7 we discuss obtained results.

\section{Types of waves in a rotating gas}

\begin{figure}[h]
\centering
\includegraphics[width=0.75\linewidth]{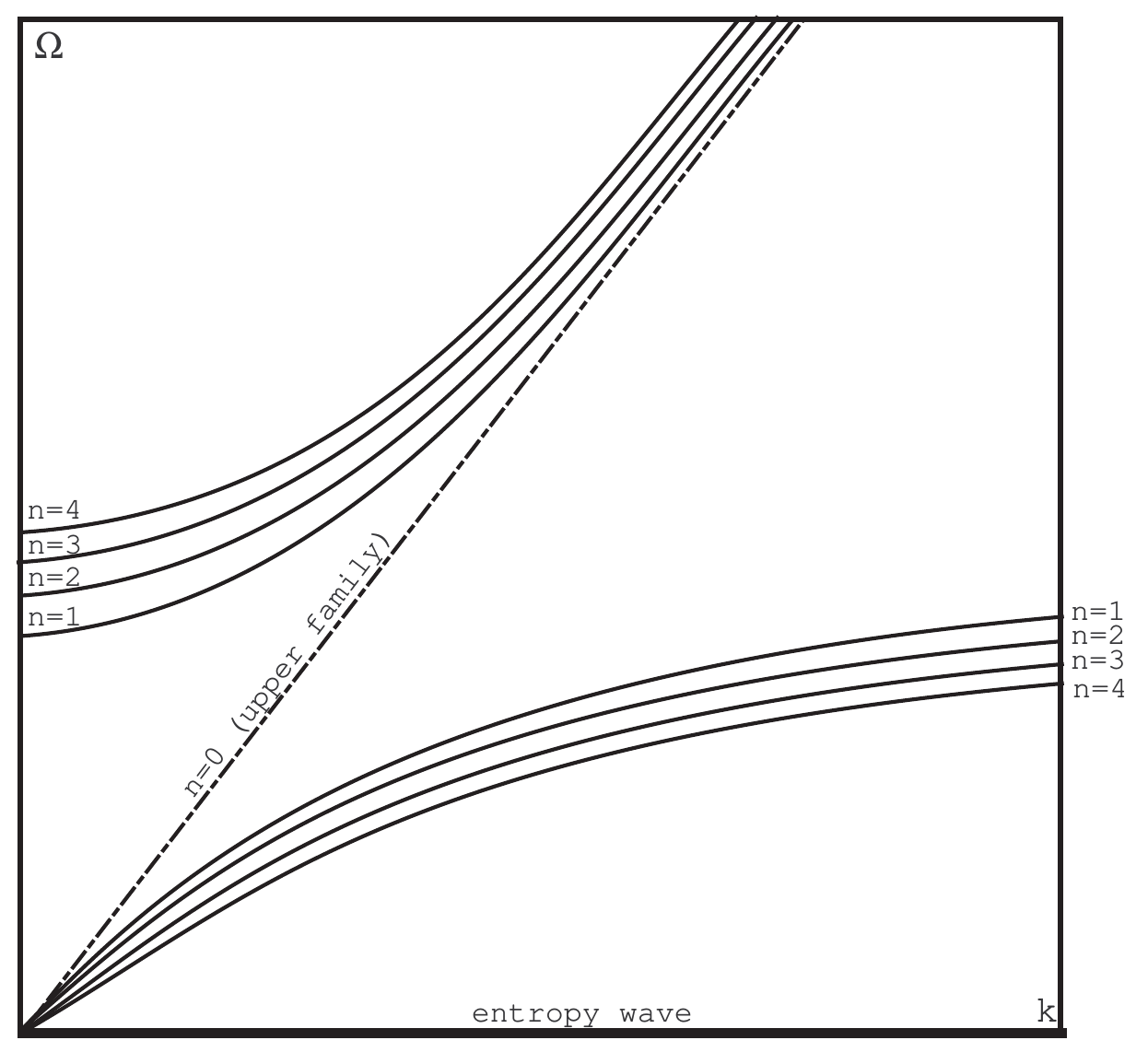}
\caption{Schematic representation  of the laws of dispersion of the waves in the rotating gas: ``upper'' family above the line corresponding to the sound wave with $\Omega=kc$, ``lower'' family  below this line and entropy wave with $\Omega=0$. }
\label{fig:families}
\end{figure}

In our previous work~\cite{GC:waves} propagation of the linear waves in ideal rotating gas has been considered. Two different families of waves with different polarizations and dispersion laws were discovered. {\bf They are defined by the equations:
\begin{equation*}
V_r = \frac{Y}{r \rho_0} e^{\frac{M \omega^2 r^2}{4 R T_0}} {,~} V_{\varphi} = \frac{ 2 \omega Y}{ i \Omega r \rho_0} e^{\frac{M \omega^2 r^2}{4 R T_0}}) {,~} V_z = \frac{ k p }{ \Omega \rho_0 } {,}
\end{equation*}
\begin{equation}
 \label{waves}
p' = \frac{1}{A} \left( \frac{ \omega Y }{ c_p T_0 } e^{\frac{M \omega^2 r^2}{4 R T_0}} - \frac{1}{r} \frac{\partial}{\partial r} \left( Y e^{\frac{M \omega^2 r^2}{4 R T_0}} \right) \right) {,~} T' = \frac{- i \Omega p' + \rho_0 \omega^2 r V_r }{ -i \rho_0 c_p \Omega } {,}
\end{equation}
where $i$ -- imaginary unit, $p'$, $T'$, $V_r$, $V_{\varphi}$, $V_z$ -- amplitude of the perturbation of the pressure, temperature, radial, azimuthal and axial velocities respectively, $\rho_0$ -- rigid body rotation density, $\Omega$ -- perturbation frequency, $k$ -- projection of the wave vector to the axial direction, $\omega$ -- angular velocity, $r$ -- cylindrical radius, $c_p$ -- gas specific heat capacity for constant pressure, $M$ -- molar mass of the gas, $R$ -- gas constant, $Y$ -- Whittaker function:
\begin{equation}
 \label{whittaker}
   Y = WM \left( \frac{B'}{4 \sqrt{-A'}}, 0.5, \sqrt{-A'} r^2 \right) {.}
\end{equation}
Here $A' = - \frac{ M^2 \omega^4 }{4 R^2 T_0^2 } + \frac{ k^2 \omega^4 }{ c_p T_0 \Omega^2 }$, $B' = - \frac{1}{\Omega^2} \left( k^2 - \frac{ \Omega^2 }{ c^2 } \right) \left( \Omega^2 - 4 \omega^2 \right)$, $c$ -- sound velocity. The equation 
$Y(r=a)=0$, where $a$ -- rotor radius, defines us the dispersion relation for the waves in the rotating gas.  They split into ``upper''  and   ``lower'' families of the waves. The ``upper'' family has degenerate case of the sound wave with disperion law
$\Omega = k c$ and 
\begin{equation*}
 V_r = 0 {,~} V_{\varphi} = 0 {,~} V_z = \frac{c}{\gamma} \frac{p'_w}{p_w} \exp \left( \frac{\left( 1 - \gamma \right) \omega^2 \left( r^2 - a^2 \right)}{2 c^2} \right) {,}
\end{equation*}
\begin{equation}
 p' = p'_w \exp \left( \frac{\omega \left( r^2 - a^2 \right)}{2 c^2} \right) {,~} T' = \frac{p'}{\rho_0 c_p} {,~} \rho ' = .....???
\end{equation}
Here $p_w$ and $p'_w$ pressure at the rotor wall and its perturbation respectively,  $\gamma$ -- adiabatic index. This case corresponds to the  zero level of  excitation of the gas in the radial direction. This wave propagates along the rotational axis and is longitudinally polarised. 

Fig.~\ref{fig:families} schematicaly represents the dispersion low of the waves in the rotating gas. 
The ``upper'' family of the waves with $\Omega \ge kc$ reduces to the conventional acoustic waves in the limit of quiescent gas. The ``lower'' family of the waves with $\Omega < kc$ reduces to the so-called ``vortex'' waves which have zero frequency and velocity of propagation in the quiescent gas~\cite{landau:hydro} (see task on pp. 315-316). These waves  have finite velocity of propagation in the rotating gas. 

Every type of waves splits into two waves propagating into opposite directions. Therefore, totally we have 4 independant waves.  }
Taking into account that the number of the waves should be equal to the number of the hydrodynamical equations, one more type of waves should exist in the rotating gas. In the limit of quiescent gas this wave should reduce to the entropy wave propagating with the zero frequency and velocity~\cite{landau:hydro}. In order to discover the last type of the waves, let us consider equations (16)-(20) from the work~\cite{GC:waves} describing dynamics of perturbations of the gas:
\begin{equation}
\label{k1}
 -i \Omega \rho' + \frac{\partial \left( r \rho_0 V_r \right) }{r \partial r} + i k \rho_0 V_z = 0 {,}
\end{equation}
\begin{equation}
\label{k2}
-i \Omega \rho_0 V_r - 2 \rho_0 \omega V_{\varphi} - \rho \omega^2 r = -\frac{\partial p'}{\partial r} {,}
\end{equation}
\begin{equation}
\label{k3}
-i \rho_0 \Omega V_{\varphi} + 2 \rho_0 \omega V_r = 0 {,}
\end{equation}
\begin{equation}
\label{k4}
 \rho_0 \Omega V_z = kp' {,}
\end{equation}
\begin{equation}
\label{k5}
 -i \rho_0 c_p \Omega T' = -i \Omega p' + \rho_0 \omega^2 r V_r {.}
\end{equation}
Here $\rho'$  is the amplitude of perturbation of  density.

Let us consider perturbations with $\Omega = 0$. In this case equations~\eqref{k1}-\eqref{k5} become
\begin{equation}
\label{k11}
 \frac{\partial \left( r \rho_0 V_r \right) }{r \partial r} + i k \rho_0 V_z = 0 {,}
\end{equation}
\begin{equation}
\label{k21}
- 2 \rho_0 \omega V_{\varphi} - \rho' \omega^2 r = -\frac{\partial p'}{\partial r} {,}
\end{equation}
\begin{equation}
\label{k31}
2 \rho_0 \omega V_r = 0 {,}
\end{equation}
\begin{equation}
\label{k41}
 kp' = 0 {,}
\end{equation}
\begin{equation}
\label{k51}
 \rho_0 \omega^2 r V_r = 0{.}
\end{equation}
{\bf Equations~\eqref{k11},~\eqref{k31}-\eqref{k51} give $V_r = 0$, $p' = 0$ and $V_z = 0$. From the equation of state
\begin{equation}
\rho_0=\frac{M}{RT_0}p_0{,} ~ \rho'=\frac{M}{RT_0} p' - \rho_0\frac{T'}{T_0} {,}
\label{rhop}
\end{equation}
we obtain
\begin{equation}
\label{k6}
\rho' = -\rho_0 \frac{T'}{T_0} {,}
\end{equation}
where $T_0$ is the  temperature of the unperturbed gas. Substitution of $p' = 0$ into equation~\eqref{k21} gives
\begin{equation}
\label{k22}
- 2 \rho_0 \omega V_{\varphi} - \rho' \omega^2 r = 0 {.}
\end{equation}
Substitution of eq.~\eqref{k6} into~eq. \eqref{k22} gives the final relationship between gas characteristics in the entropy wave:
\begin{equation}
\label{w5}
V_r = 0 {;} ~V_z = 0 {;} ~p' = 0 {;} ~{2 V_{\varphi} \over \omega r} = \frac{T'}{T_0}  {;}~ \rho' = -\rho_0 \frac{T'}{T_0} {,}
\end{equation}
with an arbitrary temperature perturbation $T'$. In such a wave Coriolis force due to the azimuthal velocity perturbation compensates an additional centrifugal force due to density perturbation. Pressure distribution remains unperturbed. It is easy to understand this result. 

From equation~\eqref{w5} one can obtain
\begin{equation}
 \label{ww5}
   \frac{2 V_{\varphi}}{\omega r} = \frac{ 2 \omega'}{\omega} = \frac{T'}{T_0} {,}
\end{equation}
\begin{equation}
 \label{www5}
   2 \ln \omega = \ln T + const {,}
\end{equation}
\begin{equation}
 \label{wwww5}
   \frac{ \omega^2 }{T} = const {.}
\end{equation}
Equation~\eqref{wwww5} means that in the  entropy wave the exponent $\exp \left( \frac{M \omega^2}{2 R T} \left( r^2 - a^2 \right) \right)$ remains unperturbed. Pressure and density satisfy to the following equations
\begin{equation}
 p = p_w \exp \left( \frac{M \omega^2}{2 R T} \left( r^2 - a^2 \right) \right), ~~ \rho = \rho_w \exp \left( \frac{M \omega^2}{2 R T} \left( r^2 - a^2 \right) \right) {,}
\end{equation}
where $\rho_w$ is the  density at the rotor wall. According to (\ref{w5}) pressure is unperturbed.  Density is perturbed due to perturbation of  $\rho_w$ like in the conventional entropy wave in accordance with equation of state for the ideal gas, see eq.~\eqref{k6}. }

So, we have totally five waves in the rotating gas: two ``upper'' family waves propagating in the opposite directions, two ``lower'' family waves also propagating in the opposite directions and the fifth wave corresponding to entropy wave (see fig.~\ref{fig:families}). The last wave has zero propagation velocity. But unlike the wave in the quiescent gas the enthropy wave perturbs the azimuthal velocity of rotation of the gas.  

Only one mode of the waves from the ``upper'' family has a damping length more than a few wavelengths. This wave has conventional dispersion relation of the form $\Omega = k c$ and propagates exactly along the axis of rotation. In this work we focus our attention on the damping of this wave only.

\section{Numerical solution of the problem}

In order to define damping of the waves in the rotating gas we have to solve hydrodynamic system of equations taking into account viscous and thermal dissipation processes. {\bf We consider an ideal gas with zero second viscosity (no excitations of the internal degrees of freedom of the molecules). The system of hydrodynamic equations describing such a gas in the cylindrical coordinate system in the axisymmetric approximation has the form~\cite{landau:hydro}:}
\begin{equation}
\frac{\partial \rho}{\partial t}+\frac{1}{r} \frac{\partial (r \rho v_r)}{\partial r} + \frac{\partial(\rho v_z)}{\partial z}=0.
\label{neprot1}
\end{equation}
\begin{eqnarray}
\rho \frac{\partial v_r}{\partial t} &+& \rho \left( v_r \frac{\partial v_r}{\partial r} + v_z\frac{\partial v_r}{\partial z}-\omega^2 r-2\omega v_{\varphi}-\frac{v^2_{\varphi}}{r} \right) = \nonumber \\
&=& -\frac{\partial p}{\partial r}+\eta \left( \left( \Delta-\frac{1}{r^2} \right) v_r+\frac{1}{3}\frac{\partial \diverg \vec{v}}{\partial r} \right) .
\end{eqnarray}
\begin{eqnarray}
\rho \frac{\partial v_{\varphi}}{\partial t} &+& \rho \left( v_r\frac{\partial v_{\varphi}}{\partial r}+v_z\frac{\partial v_{\varphi}}{\partial z}+2\omega v_r+\frac{v_{\varphi}v_r}{r} \right) = \nonumber \\
&=& \eta \left( \Delta-\frac{1}{r^2} \right) v_{\varphi}+f_{\varphi}.
\end{eqnarray}
\begin{equation}
\label{eq:vz}
\rho\frac{\partial v_z}{\partial t} + \rho \left( v_r\frac{\partial v_z}{\partial r} + v_z\frac{\partial v_z}{\partial z} \right) = -\frac{\partial p}{\partial z}+\eta \left( \Delta v_z+\frac{1}{3} \frac{\partial \diverg \vec{v}}{\partial z} \right).
\end{equation}
\begin{eqnarray}
\rho c_p \frac{\partial T}{\partial t}&+& \rho c_p \left(v_r \frac{\partial T}{\partial r} + v_z\frac{\partial T}{\partial z} \right)=\frac{\partial p}{\partial t}+v_r\frac{\partial p}{\partial r}+ \nonumber\\
& &+ v_z \frac{\partial p}{\partial z}+\lambda \left( \frac{1}{3} \frac{\partial }{\partial r}\left(r \frac{\partial T}{\partial r} \right)+\frac{\partial^2 T}{\partial z^2} \right) + \nonumber \\
+&\eta &\left( \left( \frac{\partial v_z}{\partial r}+\frac{\partial v_r}{\partial z} \right)^2 + \left( \frac{\partial v_{\varphi}}{\partial z} \right)^2+ \left( \frac{\partial v_{\varphi}}{\partial r}-\frac{v_{\varphi}}{r} \right)^2  \right. + \nonumber \\
& &+\frac{1}{2} \left( \frac{4}{3} \frac{\partial v_r}{\partial r} - \frac{2}{3} \frac{v_r}{r}- \frac{2}{3} \frac{\partial v_z}{\partial z} \right)^2 + \nonumber\\
& & + \frac{1}{2} \left( \frac{4}{3} \frac{v_r}{r} - \frac{2}{3} \frac{\partial v_r}{\partial r} - \frac{2}{3} \frac{\partial v_z}{\partial z}\right)^2+ \nonumber \\
& & \left. + \frac{1}{2} \left( \frac{4}{3} \frac{\partial v_z}{\partial z}- \frac{2}{3} \frac{\partial v_r}{\partial r}- \frac{2}{3} \frac{v_r}{r} \right)^2 \right)+q.
\label{enrot1}
\end{eqnarray}
Here $v_r, v_{\varphi}, v_z$ -- radial, azimuthal and axial velocity of the gas, $\rho$ -- density, $p$ -- pressure, $T$ -- temperature, $\eta$ -- dynamic viscosity, $\lambda$ -- thermal conductivity, $f_{\varphi}, q$ -- external force and energy source. 

{\bf The hydrodynamical approximation is valid while the path length of molecules $l_{p}$ is well below the characteristic length of the problem equal to $L\sim {RT\over M\omega a}$ or wavelength $\lambda/2\pi$. In the centrifuges these conditions are violated in the vacuum core where density of the gas is very law. 
Actually, for a correct solution of the problem it is necessary to solve the hydrodynamical  system of equations near the wall and then paste this solution with the solution in the vacuum core. This is rather difficult problems for the solution.   In this work for simplification we assume that the hydrodynamical equations are valid everywhere.}   

The conventional approach to define the damping coefficient reduces to a search of eigenvalues and eigenfunctions of the linearized system of the equations at $f=0$ and $q=0$ assuming that all the variables vary in time as $\exp(-i \Omega t)$.  The imaginary parts of the eigenvalues $\Omega$ gives us the damping coefficients of the waves. Unfortunately, it is difficult to implement this approach here because the system of the equations is too complicated. To calculate the damping coefficient we used the method of resonance.

The idea of the method is simple. Let us apply an external force $f$ with frequency $\Omega$ to the gas. Then the intensity of the excited waves $I = \frac12 \rho_0 \overline{v^2}$ will vary near the resonances according to the Lorenz equation $I/I_i \sim \delta_i ^2/((\Omega - \Omega_i)^2+\delta_i^2)$, where $\delta_i$ and $\Omega_i$ are the imaginary and real parts of the eigenvalue, $I_i$ is the intensity in the resonance. This way, the dependence on the frequency $\Omega$ of the exciting force (the resonance curve) gives us information about damping coefficient of the waves. 

For implementation of this approach it is necessary to solve the linearized system of equations for arbitrary force $f$ for a wide range of frequencies $\Omega$. The method of solution of the linearized equations for the force of the form $f \sim \sin(kz)  \sin( \Omega t)$ has been developed by us earlier in the works~\cite{GC:abramov} and~\cite{GC:verif}. 

First of all we linearize the system of equations. Similar to our previous work~\cite{GC:waves}, all the variables are presented as a sum of variables corresponding to the rigid body rotation of the gas denotes by index ``0'' and perturbation of the variables marked by upper bar. Thus,
\begin{equation}
v_r=\bar{v}_r,v_z=\bar{v}_z,v_{\varphi}=\omega r + \bar{v}_{\varphi}, p=p_0+\bar{p}, \rho=\rho_0+\bar{\rho}, T=T_0+\bar{T}.
\end{equation}

The solution of the linearized system of equations~\eqref{neprot1}-\eqref{enrot1} we search in the form
\begin{equation}
\bar{p}(r,z,t)=\left( A_1(r)+iA_2(r)\right) sin(k z) e^{-i \Omega t},
\label{sols}
\end{equation}
\begin{equation}
\bar{v}_r(r,z,t)=\left( U_1(r)+iU_2(r) \right) sin(k z)e^{-i \Omega t},
\end{equation}
\begin{equation}
\bar{v}_z(r,z,t)=(W_1(r)+iW_2(r))cos(k z)e^{-i \Omega t},
\end{equation}
\begin{equation}
\bar{v}_{\varphi}(r,z,t)=(V_1(r)+iV_2(r))sin(k z)e^{-i \Omega t},
\end{equation}
\begin{equation}
\label{eq:Tb}
\bar{T}(r,z,t)=(T_1(r)+iT_2(r))sin(k z)e^{-i \Omega t}.
\end{equation}
Substitution of the~\eqref{sols}-\eqref{eq:Tb} into~\eqref{neprot1}-\eqref{enrot1} gives the system of 10 ordinary differential equations
\begin{equation}
\frac{MgA_{2,1}}{RT_0}-\frac{\rho_0 \Omega T_{2,1} }{T_0}+\frac{(r \rho_0 U_{1,2})'}{r}-k \rho_0 W_{1,2}=0,
\label{neprrot}
\end{equation}
\begin{eqnarray}
\rho_0 \Omega U_{2,1} &-& 2\rho_0 \omega V_{1,2}-\frac{M \omega^2 A_{1,2} r}{RT_0}+\frac{\rho_0 T_{1,2} \omega^2 r}{T_0}+A_{1,2}'= \nonumber \\
&=& \eta \left( \frac{4}{3} \left( rU_{1,2} \right)''-k^2 U_{1,2}-\frac{k W_{1,2}'}{3} \right),
\end{eqnarray}
\begin{equation}
\rho_0 \Omega V_{2,1} +2\rho_0 \omega U_{1,2}=\eta \left( V_{1,2}''+\frac{V_{1,2}'}{r}-k^2 V_{1,2} -\frac{V_{1,2}}{r^2} \right)+\xi_1,
\end{equation}
\begin{equation}
\rho_0 \Omega W_{2,1} +k A_{1,2}=\eta \left( \frac{W_{1,2}'}{r}+W_{1,2}''-\frac{4 k^2 W_{1,2}}{3}+\frac{k U_{1,2}}{3r}+\frac{k U_{1,2}'}{3} \right),
\end{equation}
\begin{equation}
\rho_0 c_p \Omega T_{2,1} - \Omega A_{2,1} =\rho_0 \omega^2 r U_{1,2}+\lambda \left( \frac{T_{1,2}'}{r}+T_{1,2}''-k^2 T_{1,2} \right)+\theta_1 {.}
\label{enrot}
\end{equation}

{\bf We solve this system numerically with the following boundary conditions
\begin{equation*}
   V_r \left( 0 \right) = 0 {,~} V_r \left( a \right) = 0 {,~} V_{\varphi} \left( 0 \right) = 0 {,~} V_{\varphi} \left( a \right) = 0 {,} 
\end{equation*}
\begin{equation}
 \label{bc}
  V_z \left( a \right) = 0 {,~} T \left( a \right) = T_0 {.~} 
\end{equation}
for different frequencies $\Omega$ of the exciting force with fixed wave vector $k$ and obtain the resonance curve. Approximation of this curve by the Lorenz profile gives us an imaginary part of the frequency of the waves and, therefore, damping coefficient. Accuracy of the method of resonance depends on the  number and location of peaks on the resonance curve. We approximated the line profile in all the frequency range. 
Our approach gives satisfactory results while the width of the line $\Delta \Omega$ is much less the distance 
between the peaks. }

\section{Verification of the method}

\subsection{Analytical solution}

For verification of the method we consider an exactly solvable case of the wave damping in the quiescent gas. In this case the rotor does not rotate and gas pressure and density are constant. Damping of the waves occurs due to two mechanisms. Firstly, damping of the waves is defined by the viscous stresses and heat exchange between parts of the gas itself (volume damping mechanism). Secondly, the damping of the waves occurs due to viscous stresses and heat exchange  with the walls of the rotor (surface damping mechanism). The volume damping coefficient is as follows~\cite{landau:hydro}
\begin{equation}
 \gamma_1=\frac{\Omega^2}{2 \rho c^3}\left( \frac{4}{3}\eta +\lambda\left( \frac{1}{c_v}-\frac{1}{c_p}\right)\right) {,}
\label{g1}
\end{equation}
where $\rho$ -- density of the gas, $c$ -- sound velocity $c_v$ -- the specific heat of the gas at constant volume. The surface damping coefficient is~\cite{landau:hydro}
\begin{equation}
 \gamma_2=\frac{\sqrt{\Omega}}{\sqrt{2} a c}\left( \sqrt{\frac{\eta}{\rho}}+\left( \frac{c_p}{c_v}-1\right)\sqrt{\frac{\lambda}{\rho c_p}}\right) {.}
\label{g2}
\end{equation}

For verification of the method we consider only volume damping. The system of equations~\eqref{neprot1}-\eqref{enrot1} at $\omega=0$ can be rewritten as:
\begin{equation}
 \label{eq:es1}
 \frac{\partial \rho}{\partial t}+\rho_0 \frac{\partial v_z}{\partial z}=0,
\end{equation}
\begin{equation}
 \label{eq:es2}
 \rho_0 \frac{\partial v_z}{\partial t}=-\frac{\partial p}{\partial z}+\eta \frac{4}{3}\frac{\partial^2 v_z}{\partial z^2}+f_z,
\end{equation}
\begin{equation}
 \label{eq:es3}
 \rho_0 c_p\frac{\partial T}{\partial t}=\frac{\partial p}{\partial t} {.}
\end{equation}

We take the driving force in the following form $f_z=F_z(t) \sin(kz)$ and look for a solution in the following form: 
\begin{eqnarray}
 & & \bar{p}(z,t)=p'(t)\cos(kz), ~ \bar{\rho}(z,t)=\rho'(t)\cos(kz), \nonumber \\
 & & \bar{v}_z(z,t)=V_z(t)\sin(kz), ~ \bar{T}(z,t)=T'(t)\cos(kz) {.}
\end{eqnarray}

The system of the equations~\eqref{eq:es1}-\eqref{eq:es3} can be rewritten as
\begin{equation}
 \frac{\partial \rho'}{\partial t}+\rho_0 k V_z=0,
\label{nepr2}
\end{equation}
\begin{equation}
 \rho_0 \frac{\partial V_z}{\partial t}=k p' - \frac{4}{3}\eta k^2 V_z + F_z(t),
\label{ns2}
\end{equation}
\begin{equation}
 \rho_0 c_p\frac{\partial T'}{\partial t}=\frac{\partial p'}{\partial t} {.}
\label{en2}
\end{equation}
We substitute $\rho'$ from equation~\eqref{rhop} into equation~\eqref{nepr2} and express $p'$ from equation~\eqref{ns2}:
\begin{equation}
 \frac{M}{RT_0}\frac{\partial \bar{p}}{ \partial t}-\frac{\rho_0}{T_0}\frac{ \partial T'}{ \partial t}+\rho_0 k V_z=0,
\label{nepr3}
\end{equation}
\begin{equation}
p' = \frac{\rho_0}{k}\frac{\partial V_z}{\partial t}+\frac{4}{3}\eta k V_z-\frac{F_z}{k},
\end{equation}
\begin{equation}
 \frac{\partial T'}{\partial t}=\frac{1}{\rho_0 c_p}\frac{\partial p'}{\partial t}.
\label{en3}
\end{equation}
Substitution of eq.~\eqref{en3} into eq.~\eqref{nepr3} gives
\begin{equation}
 \left(\frac{M}{RT_0}-\frac{1}{c_p T_0}\right)\frac{\partial p'}{ \partial t}+\rho_0 k V_z=0,
\label{nepr4}
\end{equation}
\begin{equation}
p' = \frac{\rho_0}{k}\frac{\partial V_z}{\partial t}+\frac{4}{3}\eta k V_z-\frac{F_z}{k}{.}
\label{ns4}
\end{equation}
Since $\frac{M}{RT_0}-\frac{1}{c_p T_0}=\frac{1}{c^2}$, eq.~\eqref{nepr4} can be rewritten as:
\begin{equation}
 \frac{1}{c^2}\frac{\partial p'}{ \partial t}+\rho_0 k V_z=0{.}
\label{nepr5}
\end{equation}
Substitution of eq.~\eqref{ns4} into eq.~\eqref{nepr5} gives the second order differential equation:
\begin{equation}
 \rho_0 \frac{\partial^2 V_z}{\partial t^2}+\frac{4}{3}\eta k^2 \frac{\partial V_z}{\partial t}+\rho_0 k^2 c^2 V_z=\frac{\partial F_z}{\partial t}.
\label{maineqn}
\end{equation}
This is the equation of the induced damped oscillations. Substitution of $F_z=F_z^0 \sin \left( \Omega t \right)$ gives the solution of this equation at $t\rightarrow \infty$ in the form:
\begin{equation}
 V_z=\frac{F_z^0 \Omega}{\rho_0 \sqrt{\left( \Omega_0^2 - \Omega^2 \right)^2 + \frac{16}{9}\frac{\eta^2 k^4 \Omega^2}{\rho^2_0}}}\cos \left( \Omega t +\phi \right),
\end{equation}
where
\begin{equation}
 \phi = \arctan \left( \frac{4}{3} \frac{\eta k^2 \Omega}{\rho_0 \left( \Omega_0^2 - \Omega^2 \right)}\right).
\end{equation}
and $\Omega_0 = k c$ -- resonance frequency of the system.

Intensity of the oscillation is defined by the expression:
\begin{equation}
  I=\frac{\rho_0 \overline{V^2_z}}{2}=\frac{\left( F_z^0 \Omega \right)^2}{ \left( \Omega_0^2 - v^2\right)^2+\frac{16}{9}\frac{\eta^2 k^4 \Omega^2}{\rho^2_0}}.
\label{maine}
\end{equation}
If we introduce the following notation $\sigma^2=\frac{16}{9}\frac{\eta^2 k^4}{\rho^2_0}$, the damping coefficient of the wave can be written as 
\begin{equation}
 \gamma_1=\frac{\sigma}{2 c},
\label{gammasigma}  
\end{equation}
where $\sigma$ is the width of the resonance curve~\eqref{maine}.

\begin{figure}[h]
\centering
\includegraphics[width=0.9\linewidth]{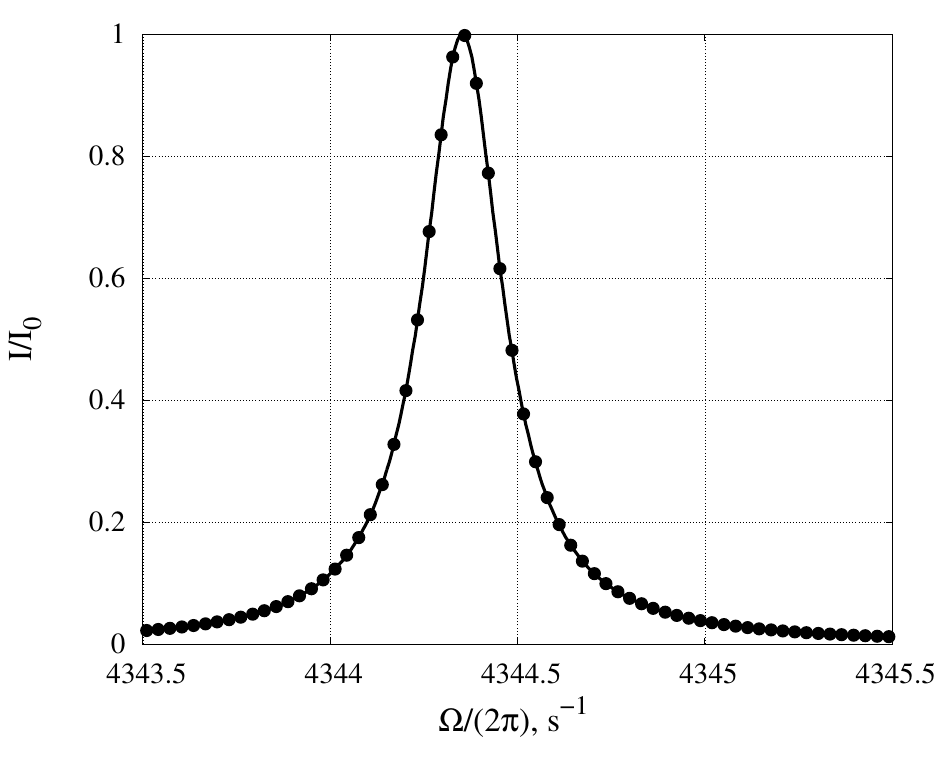}
\caption{Dependence of the intensity of waves on the frequency of the exciting force for wave vector $\frac{k}{2\pi}=50 ~ m^{-1}$. Solid line -- analytical calculation according eq. ({\protect\ref{maine}}), Solid circles  -- numerical results}
\label{fig:spektr1}
\end{figure}

\begin{figure}[h]
\centering
\includegraphics[width=0.9\linewidth]{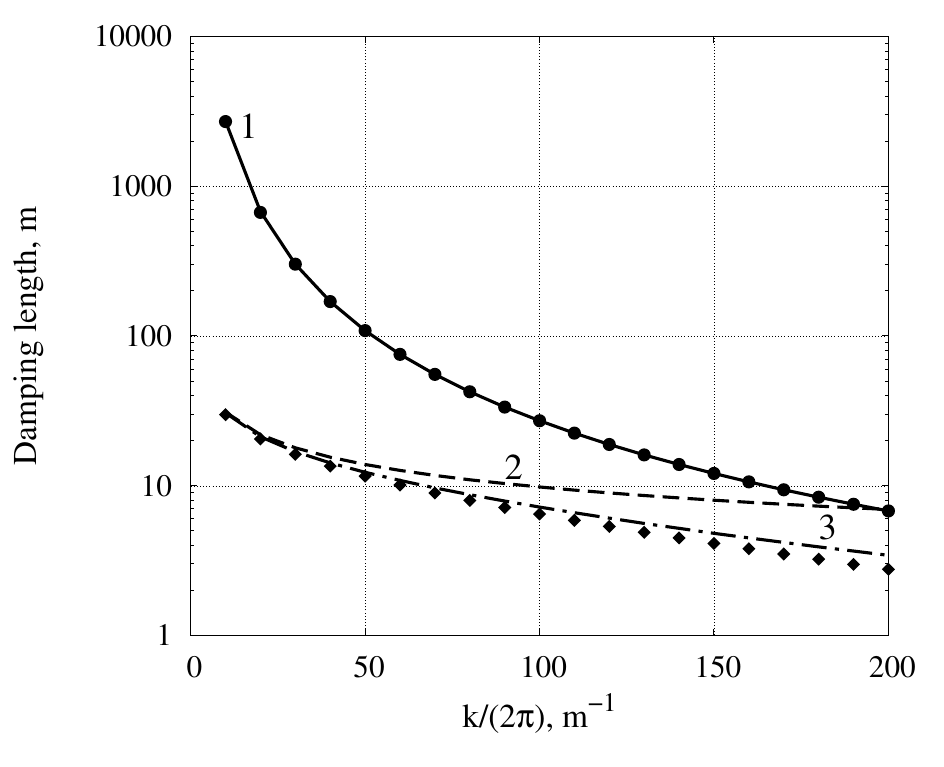}
\caption{Damping length of acoustic waves versus wave vector of the driving force (for the quiescent gas). Solid line 1 -- analytical dependence (\ref{g1}), dashed line 2 -- analytical dependence (\ref{g2}), dashdot line 3 -- volume and surface damping, \textbullet -- numerical results for volume damping, $\blacklozenge$ -- numerical results for volume and surface damping}
\label{fig:dampw0}
\end{figure}

\subsection{Comparison of the analytical and numerical solutions}

For the verification of the proposed method, the resonance curve has been calculated numerically by solution of the set of equations~\eqref{neprrot}-\eqref{enrot} at $\omega=0$ (no rotation). The parameters of the rotor and the gas used at the calculations are given in tab.~\ref{tab1}. Boundary conditions at the rotor walls were specified as an adiabatic free slip (no friction) walls to avoid damping of the waves at the walls. Calculations were performed for $UF_6$ at temperature $T=300 ~ K$. Comparison of the analytical equation~\eqref{maine} with the numerical calculations is shown in fig. \ref{fig:spektr1}. The numerical points agree with the analytical curve with very high accuracy (less than 1\%). 

Estimation of the width of the resonance curve $\sigma$ were obtained by fitting of the numerical results by the Lorentz profile in the form of~\eqref{maine}. The fitting was performed by the $\chi$-squared method and the damping length was defined as 
\begin{equation}
 L=\frac{2c}{\sigma}.
\end{equation}
The dependence of the damping length on the wave-vector is shown in fig.~\ref{fig:dampw0} for two cases. The first case corresponds to adiabatic free slip walls of the rotor. In this case no damping of the waves due to the friction and heat exchange at the walls takes place. Comparison shows that the damping length defined by our method well agrees with the dependence defined by eq.~\eqref{g1}.

In the second case calculations were performed assuming no slip wall of the rotor having constant temperature. In this case the damping of the waves occurs not only due to the velocity and temperature gradients in the volume of the gas (volume damping) but also due to the velocity and temperature gradients which arise due to viscous friction and heat exchange of the gas with the wall (surface damping).  The damping rate due to interaction of the wave with the wall is defined by eq.~\eqref{g2}. According to fig.~\ref{fig:dampw0}, the damping of the waves occurs mostly due to interaction of the wave with the rotor wall for the typical parameters of the Iguassu gas centrifuge.

The total damping rate coefficient $\gamma_{tot}$ can be estimated as a sum $\gamma_{tot}=\gamma_1+\gamma_2$ of the volume and surface damping coefficients. {bf\ This equation will be justufied below after eq. (\ref{emec2})}. The damping length is defined as $L_{tot}={1/ \gamma_{tot}}$. $L_{tot}$ is shown on fig.~\ref{fig:dampw0} by rhombuses. The discrepancy between the numerical and analytical results for the total (volume and surface) damping is within 5~\% at $k/2\pi < 50$ $m^{-1}$ and about 15 $\%$ at $k=200$ $m^{-1}$. This difference is not important for us because characteristic frequency of the rotation is {\bf $\omega \sim 2 \pi \cdot 1700$~$s^{-1}$ (see tab.~\ref{tab1}) gives the value of the characteristic wave vector ($k/2\pi \sim 20$ $m^{-1}$).  The deviation of the damping coefficient from  simple sum of the volume and surface damping coefficients becomes significant at high wave vectors.  The volume damping agrees with the predictions with high accuracy. Apparently, the interaction of the wave with the cylindrical surface of the rotor has more complicated character than it folows from eq. (\ref{g2}). Unfortunately, an accurate solution of the problem of damping of the sound wave in tube is absent. Nevertheless, we consider the results of verification as succesful because  in the range of the  wave length interesting for us the agreement of the results obtained with the resonance method and with theoreitical estimates is  satisfactory.}

\section{Damping of the waves in strong centrifugal field}

\begin{figure}[ht]
\centering
\includegraphics[width=0.8\linewidth]{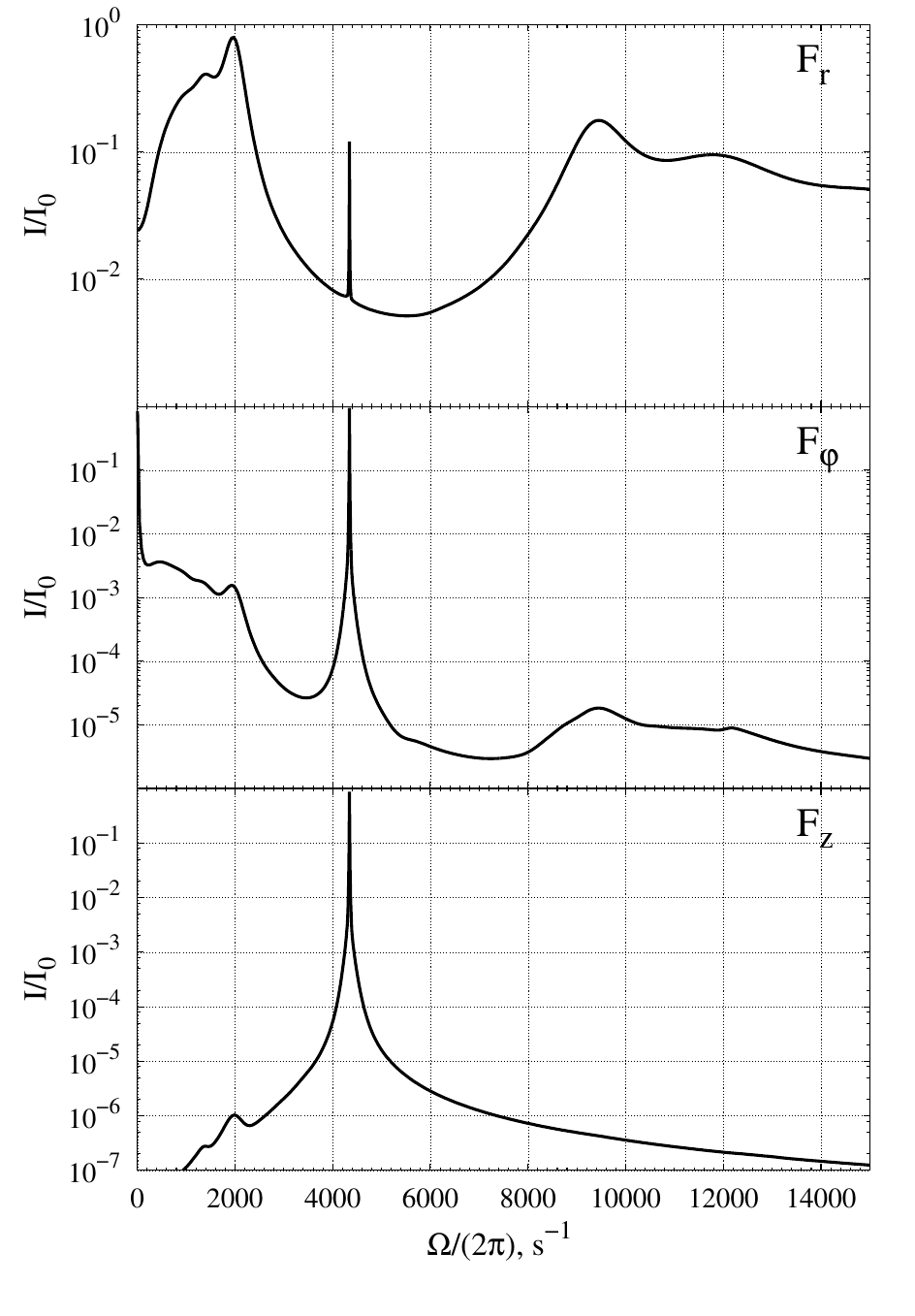}
\caption{Resonance curves for different polarizations of the exciting force $F_r,~F_{\varphi},~F_z$ for the gas in the rotating rotor with parameters from table {\protect\ref{tab1}.} } 
\label{fig:w1700sp}
\end{figure}

The developed method has been applied for the gas in strong centrifugal field. Resonance curves obtained for different polarizations of the exciting force with $\frac{k}{2 \pi} = 51 ~m^{-1}$ are shown in fig.~\ref{fig:w1700sp}. Calculations were performed for parameters of the rotor given in tab.~\ref{tab1}. For all cases there is very narrow resonance at the frequency $4385 \rm ~ Hz$ when the force is polarized along the rotational axis. This resonance corresponds to the acoustic waves. Another narrow resonance is located at $\Omega =0$ (see the central panel in fig.~\ref{fig:w1700sp}). This resonance corresponds to the entropy wave discussed in sec. 2. There are also a wide resonances at frequencies $2000 \rm ~ Hz$ and $9300 \rm ~ Hz$ when the exciting force is polarized in radial and azimuthal directions. They correspond to the ``lower'' and ``upper'' families of the waves in the gas in strong centrifugal field predicted in our previous paper~\cite{GC:waves}. The width of these two resonances is comparable with the resonance frequency. This means that these waves decay on the scale comparable with the wavelengths. In this work we focus our attention on the damping of the acoustic waves because they have the largest damping length. 

\begin{table}
\caption{\label{tab1}Parameters of the rotor and gas}
\begin{center}
\begin{tabular}{cc}

Parameter & Value  \\

$M$ & 352 g/mol\\
$a$ & 65 mm \\
$T_0$ & 300 K\\
$p_0$ & 80 mm Hg\\
$c_p$& 385 $\rm J/kg \cdotp K$ \\
$c$& 86 m/s\\
$\eta$ & $1.83*10^{-5}$ $\rm Pa \cdotp s$\\
$\lambda$ & 0.0061 $J/(m \cdotp s \cdotp K)$\\
$\omega$ & $2 \pi \cdotp 1700 ~ s^{-1}$\\
\end{tabular}
\end{center}
\end{table}

\begin{figure}[ht]
\centering
\includegraphics[width=0.9\linewidth]{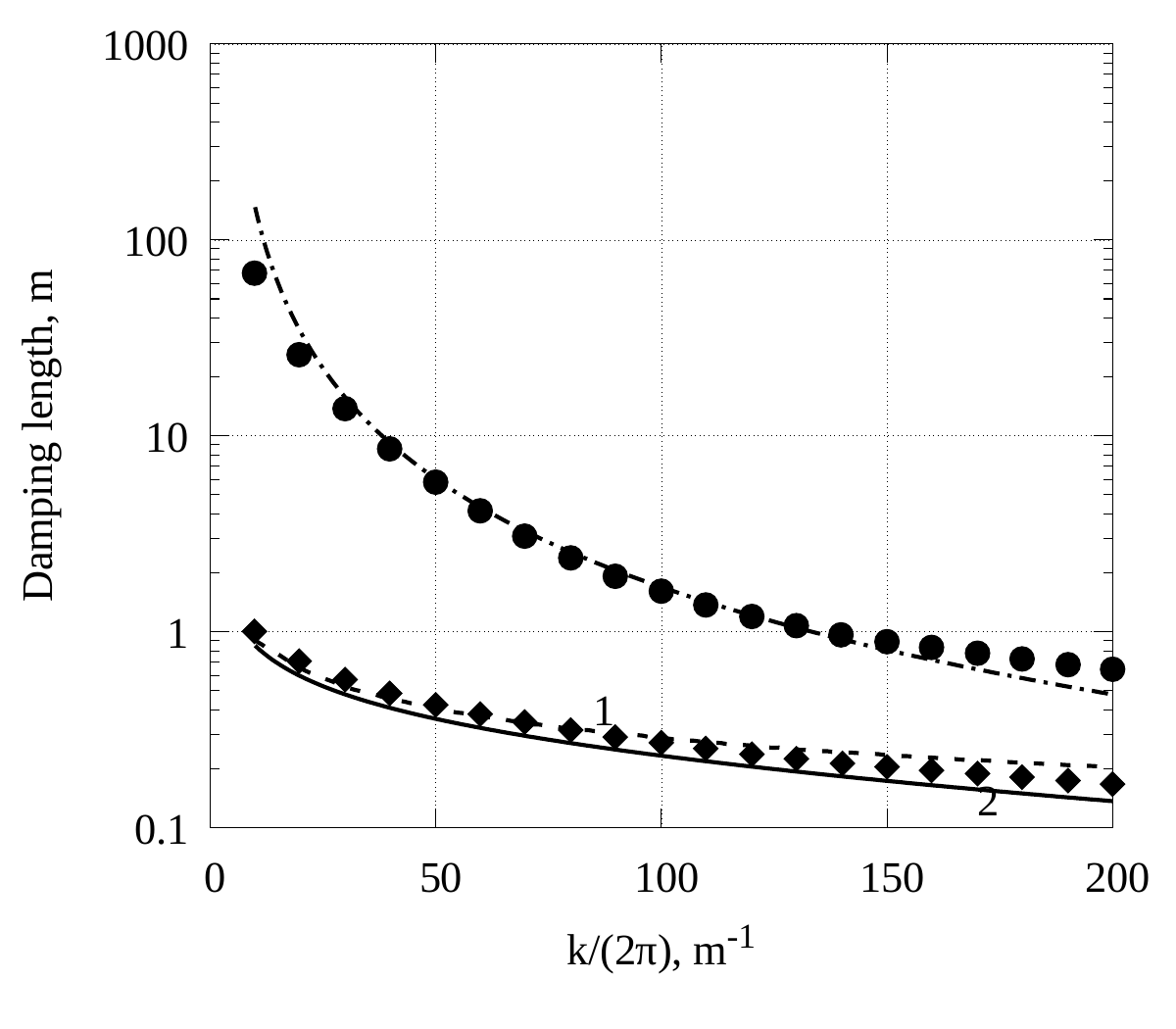}
\caption{Damping length versus wave length. \textbullet -- numerical calculations for slip adiabatic walls of the rotor (volume damping only), $\blacklozenge$ -- numerical calculations for full damping (no slip walls), dashdot line -- analytical estimation of the volume damping~\eqref{gamma_vol}, dashed line 1 -- analytical estimation of the surface damping~\eqref{gammarot}, solid line 2 -- sum of the analytical estimations of the surface and volume dampings}
\label{fig:w1700damp}
\end{figure}

Damping length of the acoustic waves calculated from the numerical solution of eqs.~\eqref{neprrot}-\eqref{enrot} in rotating gas is shown in fig.~\ref{fig:w1700damp} by circles for volume damping and rhombuses for total (surface and volume) damping. It follows from fig.~\ref{fig:w1700damp} that the surface damping of the acoustic waves in the rotating gas dominates the volume damping and almost all energy loss is due to friction and heat exchange with the rotor wall. 

\section{Analytical estimation of the damping}

According to~\cite{landau:hydro}, the damping coefficient $\gamma_d$ of a wave can be obtained as 
\begin{equation}
 \label{gammad}
 \gamma_d=\frac{|\dot{E}_{mec}|}{2 E c},
\end{equation}
where $\dot{E}_{mec}$ -- rate of the mechanical energy dissipation, $E$ -- average energy of the wave. Rate of the mechanical energy dissipation is equal to~\cite{landau:hydro}
\begin{equation}
 \label{eq:emec}
 \dot{E}_{mec}=-\frac{\lambda}{T} \int \overline{\left( \nabla T \right)^2} dV -\frac{\eta}{2} \int \overline{\left( \frac{\partial v_i}{\partial x_k}+\frac{\partial v_k}{\partial x_i}-\frac{2}{3}\delta_{ik}\frac{\partial v_l}{\partial x_l} \right)^2} dV,
\end{equation}
and
\begin{equation}
\label{eq:E}
 E=\int \left( \frac{\rho_0 \overline{v^2}}{2} + \frac{c^2 \overline{\rho'^2}}{2 \rho_0} \right) dV.
\end{equation}

Energy dissipation~\eqref{eq:emec} is determined by velocity and temperature gradients. Average energy in the wave can be calculated by substitution of the axial velocity profile for dissipationless rotating gas (equation (54) from the work~\cite{GC:waves}) 
\begin{equation}
\label{eq54}
 v_z = v_0 \exp \left( \frac{(1 - \gamma) \omega^2 (r^2 - a^2)}{2c^2} \right) e^{i \Omega t} {,}
\end{equation}
where $v_0 = \frac{c}{\gamma} \frac{\bar{p}_w}{p_w}$ -- amplitude of velocity in the wave, and density profile 
\begin{equation}
\rho = \rho_w \exp \left( \frac{\gamma \omega^2 (r^2 - a^2)}{2c^2} \right)  {,}
\end{equation}
into integral~\eqref{eq:E}. Then we obtain
\begin{eqnarray}
 E=2 \pi L \int^a_0 \frac{\rho_w e^{A \left(\frac{r^2}{a^2} - 1 \right)} v^2}{2} r dr = \nonumber \\
= 2 \pi L a^2 \frac{\rho_w v^2_0}{2 A} \frac{\gamma}{2 - \gamma} \left( 1 - \exp \left( -A \frac{\left( 2 - \gamma \right)}{\gamma} \right) \right) = \nonumber \\
= \frac{E_0}{A} \frac{\gamma}{2 - \gamma} \left( 1 - \exp \left( -A \frac{\left( 2 - \gamma \right)}{\gamma} \right) \right) {,}
 \label{energy}
\end{eqnarray}
where $A = \frac{M \omega^2 a^2}{2 R T_0}$, $E_0$ -- energy of the wave with amplitude $v_0$ in uniform 
gas with density $\rho_w$.

Calculation of  integral \eqref{eq:emec} can be divided on two parts corresponding to the volume and surface damping as follows
\begin{equation}
  \dot{E}_{mec}= \int_{r<a-h}(...) dV + \int_{a-h <r < a} (...) dV,
  \label{emec2}
\end{equation}
where $h$ is the thickness of layer where the amplitude of the plane wave decays at approaching to the wall due to friction and heat exchange with the wall. 
 The characteristic thickness of this layer is~\cite{landau:hydro}:
\begin{equation}
 h \approx \sqrt{\frac{2 \nu}{\Omega}},
\end{equation}
where $\nu$ -- kinematic viscosity, $\Omega$ -- wave frequency. 
{\bf Eq. (\ref{emec2}) shows that the damping koefficient can be presented as the sum the volume and surface damping provided that $h \ll a$. }

\subsection{Surface damping}

In the region $ a-h < r < a$ the perturbation of velocity and temperature go to zero due to friction and heat exchange with the wall. At the condition 
\begin{equation}
\label{eq:cond_surf}
 h=\sqrt{\frac{2 \nu}{\Omega}} \ll \frac{a}{A} {.}
\end{equation}
the thickness of the surface layer is much less the characteristic scale of variation of density and pressure in the gas. In this case we can neglect variation of these values at the calculation of the surface integral in eq. \eqref{emec2}.  Then, this integral $\dot E_{mec}^{surf}$ 
exactly coincides with the similar integral for ordinary plane wave interacting with the wall and damping coefficient equals
\begin{equation}
 \gamma_d^{surf}=\frac{|\dot{E}_{mech}^{surf}|}{2 E c}=A {(2-\gamma)\over \gamma} {|\dot E_{mec}^{surf}|\over 2E_0 c} \left( 1 - \exp \left( -A \frac{\left( 2 - \gamma \right)}{\gamma} \right) \right).
\label{gammarot}
\end{equation}
But 
\begin{equation}
 {|\dot E_{mec}^{surf}| \over 2E_0 c}=\gamma_2.
\end{equation}
Therefore, the surface damping in the rotating gas equals
\begin{equation}
\gamma_d^{surf} = \gamma_2 A \frac{\left( 2 - \gamma \right)}{\gamma} \left( 1 - \exp \left( -A \frac{\left( 2 - \gamma \right)}{\gamma} \right) \right) {.}
\end{equation}
 
\subsection{Volume damping}

For the calculation of $\dot{E}_{mec}$ in the volume it is necessary to have correct solution for the velocity and temperature variation in the wave. Exploration of the solution obtained in the dissipationless approximation gives strongly overestimated damping. Unfortunately, we have no analytical solution of the problem. Therefore, we give approximate estimate of the volume damping. Integral in~\eqref{eq:emec} is accumulated in all the volume of the rotor. Viscosity and thermal conductivity strongly suppress the velocity of the wave reducing the amplitude to zero below radius $a_{min}$ where the viscous terms in the equations dominate over the inertial terms. According to eq.~\eqref{eq:vz} this happens when 
\begin{equation}
 \rho c = \eta k.
\end{equation}
This gives equation for $a_{min}$ in the form
\begin{equation}
 \label{eq:amin}
 c \rho_w \exp \left( \frac{M \omega^2 a^2}{2 R T_0} \left( \frac{a^2_{min}}{a^2} - 1 \right) \right) = k \eta {.}
\end{equation}
In the interval between $a_{min}$ and $a$ the wave velocity is described by eq.~\eqref{eq54}. The temperature perturbation is also described by solution from~\cite{GC:waves}. These equations do not take into account the interaction with the wall. Therefore the integration on $r$ can be extended from $a-h$ to $a$ because $h \ll a$. Then,
integrating  ~\eqref{eq:emec} in the interval from $a_{min}$ to $a$ we obtain the following equation for $\dot{E}_{mec}$
\begin{eqnarray}
 & \dot{E}_{mec}^{vol} = - \frac{\pi L}{2} \frac{\bar{p}_w^2}{\rho_w^2} \left( \left( \frac{\lambda}{c_p^2 T_0} + \frac{\eta}{c^2} \right) \left( 2 A \Gamma + 1 + \frac{1}{A} \left( \frac{k \eta}{c \rho_w} \right)^{- 2 \Gamma}  \ln \frac{k \eta}{c \rho_w} + \right. \right. \nonumber \\
 & + \left. k^2 \frac{a^2}{2 A \Gamma} \left( \frac{\lambda}{c_p^2 T_0} + \frac{4}{3} \frac{\eta}{c^2} \right) \left( 1 - \left( \frac{k \eta}{c \rho_w} \right)^{- 2 \Gamma} \right) \right) {,}
 \label{emec_vol}
\end{eqnarray}
where $\Gamma = \frac{\gamma - 1}{\gamma}$. Energy of the wave $E$ has been calculated according to eq.\eqref{energy}. Then, 
volume damping coefficient equals
\begin{eqnarray}
 & \gamma_d^{vol}=\frac{|\dot{E}_{mech}^{vol}|}{2 E c} = \nonumber \\
 & = \frac{A}{a^2} \frac{2 - \gamma}{\gamma} \left( 1 - \exp \left( - A \frac{ \left( 2 - \gamma \right)}{\gamma} \right) \right)^{-1} \cdot \nonumber \\
 & \cdot \left( \left( \frac{\lambda}{c_p^2 T_0} + \frac{\eta}{c^2} \right) \left( 2 A \Gamma + 1 + \frac{1}{A} \left( \frac{k \eta}{c \rho_w} \right)^{- 2 \Gamma}  \ln \frac{k \eta}{c \rho_w} + \right. \right. \nonumber \\
 & + \left. k^2 \frac{a^2}{2 A \Gamma} \left( \frac{\lambda}{c_p^2 T_0} + \frac{4}{3} \frac{\eta}{c^2} \right) \left( 1 - \left( \frac{k \eta}{c \rho_w} \right)^{- 2 \Gamma} \right) \right) {.}
\label{gamma_vol}
\end{eqnarray}

In fig.~\ref{fig:w1700damp} analytical estimation calculated from~\eqref{gamma_vol} is shown by dash-dotted line. The analytical approximation agree with the numerical results in the limit of 20\% at $\frac{k}{2 \pi} \sim 20 ~ m^{-1}$. Taking into account that the volume damping 2 orders of magnitude less than surface damping, this discrepancy gives negligible (below 1\%) error in the total damping.

\section{Conclusions}

{\bf  There is the  mode of the waves propagating exactly along the rotor and polarized along the axis of rotation in the gas rotating in the GC. 
We call it an acoustic (or sound) wave. The wave have the law of dispersion similar to the conventional sound waves in the quiescent gas. 
 In the result of the work  we propose analytical equations defining the damping length of the acoustic wave in dependence on the parameters of the rotor and working gas. The problem is solved in the hydrodynamical approximation valid if  the path length of molecules is much less the characteristic scale of variation of the 
hydrodynamical variables: speed, density, pressure etc.  Typically, this is not   fulfilled in all volume of  real gas centrifuges. They have a vacuum core near the rotation axis where 
the hydrodynamics can not be applied. This problem is typically avoided by considering the gas dynamics in the layer located close to the wall of the rotor and imposing boundary conditions at the fictitious internal boundary located in the region where the Knudsen number $\sim 1$ \cite{}. Experience shows that if to  continue the solution to the rotational axis, the solutions practically does not differ. Apparently this happens because there is no mass in the vacuum core. Therefore it does not matter what approximation is used for description of the gas with mass close to zero. 

Comparison of the results obtained by the method of resonances with theoretical predictions shows excellent agreement for the volume damping in the quiescent gas. But 
for the total damping we have visible disagreement at high frequencies which achieve 15\% at $k/2\pi \sim 200$.  The same picture takes place in the case of rotating  gas. We guess that  apparently interaction of the wave with the rotor wall has more complicated character than it is given by eq. (\ref{g2}).  Surprisingly, we do not found in literature an accurate solution of the problem for damping of the waves in the tube filled by the  quiescent gas. The question remains open. 
In spite of these uncertainties,  the obtained equations provide accuracy sufficient for practical estimates. 

It follows from these equations that  the acoustic wave propagates along the Iguassu GC almost without damping. In the considered case the damping length $\sim$0.7 m, while the length of the Iguassu GC is 0.48 m. 
According to the equations, increase of the length of the rotor does not change this situation because the optimal pressure of GC is proportional to the rotor length~\cite{GC:Wood}. In all well optimized GC the waves are able to propagate from one  to  another end of the rotor. The role  of the waves becomes even more interesting at the increase of the rotor velocity $V$. According to our results the damping length decreases with the rotor velocity as $V^{-2}$ while the optimized pressure increases as  $V^5$~\cite{GC:Wood}. Therefore, the  faster the rotor rotates, the larger the damping length in the well optimized GC. Thus, our results show that the impact of the waves in industrial GC can be especially important at the exploration of GC with fast rotating rotor and this effect should be taken into account at their design.}

\begin{acknowledgements}
The present work was supported by Russian science  foundation, project N~18-19-00447. 
\end{acknowledgements}

\newpage
\bibliographystyle{spmpsci}

\end{document}